# A Task-Agnostic Algebraic Integrity Metric for Event-Camera Streams Toward SOTIF-Compliant Perception using Pearson Correlation Coefficient

Arthur de Miranda Neto

*Federal University of Lavras (UFLA), Brazil*

arthur.miranda@ufla.br

**Abstract —** Event cameras have emerged as a high-bandwidth, low-latency sensing modality for safety-critical perception in automated driving systems (ADS), offering microsecond temporal resolution, 120–140 dB dynamic range, and intrinsic absence of motion blur. However, no task-agnostic quality metric currently operates directly on the asynchronous event stream: state-of-the-art proxies require a downstream task (e.g., detection accuracy, tracking error) to assess stream integrity, which is incompatible with the certification requirements of ISO 21448 (SOTIF) and ISO/PAS 8800:2024. The recent BiasBench benchmark (CVPR 2025) explicitly identifies this gap. This work proposes a unified algebraic framework that lifts the Pearson Correlation Coefficient (PCC) — historically used in two prior works for redundancy filtering and ROI selection on frame-based images, to the three standard event representations: Time Surface, Event Frame, and Voxel Grid. The framework yields three metrics: (i) *r-TS* for stream integrity monitoring against an ego-motion-predicted Time Surface, (ii) $r_2$*-EF* for adaptive ROI selection requiring only integer comparisons, and (iii) *r-VG* for temporal redundancy gating. A structural isomorphism is established between the contrast-threshold mechanism of the event camera ($|\Delta L| \geq C$) and the PCC-based change criterion, the three lifted metrics are formalized, and pipeline latency and information loss are analyzed symmetrically against the raw stream. Illustrative behavior of each metric is demonstrated on a procedural-synthetic event stream, explicitly generated by direct simulation of the emission model rather than drawn from any real or video-derived dataset, including a tunnel-dip integrity-anomaly scenario in which $r_C$ drops from 0.93 (coherent flow) to below 0 (alarm). This work adopts an explicit epistemic convention, [ESTABLISHED], [SOLID], [HYPOTH.], [OPEN] — to delineate the status of every contribution.

**Epistemic-status convention applied to every contribution. [ESTABLISHED]** Result drawn from published work. **[SOLID]** Extension technically grounded in the literature, defensible without new experimental data, to be validated. **[HYPOTH.]** Original research hypothesis, formally motivated but not yet tested. **[OPEN]** Scientific lock acknowledged as open in literature. **[ILLUSTR.-SYNTH.]** Quantitative result computed on the procedural-synthetic stream defined in §6.1; not a result on real or video-derived data.

## 1. Introduction

The dependability of embedded perception in automated driving systems (ADS) has rested for three decades on a single paradigm: the fixed-rate frame camera. This paradigm reaches structural limits in three families of safety-critical scenarios: (i) tunnels and back-lit conditions, where the ~60 dB dynamic range of CMOS sensors saturates; (ii) high-dynamic-motion situations, where the 10–33 ms inter-frame interval exceeds the reaction window of automated emergency braking; and (iii) rapid glare events that produce irreducible motion blur.

Event cameras [1, 2] — bio-inspired neuromorphic sensors, represent a paradigmatic break. Each pixel operates asynchronously and independently, emitting an event e = (x, y, t, p) only when a local log-luminance variation exceeds a configurable contrast threshold C. The result is microsecond-level temporal resolution, 120–140 dB dynamic range, structural absence of motion blur, and an order-of-magnitude reduction in power consumption. Industrial maturity is now confirmed: Sony Semiconductor (IMX636/637), Prophesee (Metavision EVK4, OpenEB), iniVation (DAVIS346, Speck), and SynSense (DYNAP-CNN) deploy neuromorphic sensors and chips in automotive applications. Gehrig and Scaramuzza [3] recently demonstrated, in Nature (2024), that a hybrid event/frame system attains the latency equivalent of a 5,000-fps camera with the bandwidth of a 45-fps camera.

Yet, despite this hardware maturity, a fundamental gap remains for safety certification. ISO 21448 (Safety of the Intended Functionality, SOTIF) [4] and the recent ISO/PAS 8800:2024 [5] require quality and integrity metrics whose validity does not depend on a downstream applicative task. As of 2025, the BiasBench benchmark [6] explicitly states that no metric of event-stream quality operating directly on the (x, y, t, p) flow, without dependence on a downstream task, exists in the literature. SpikeSlicer [7] adapts the temporal segmentation, but its training signal still depends on tracking or recognition task.

This paper proposes a PCC-Event, a unified algebraic framework that fills this gap. The core idea is a bridge between two prior works on frame-based perception, a global redundancy criterion using the Pearson Correlation Coefficient (PCC) [8] and an automatic ROI-selection mechanism using a binarized PCC [10] and the asynchronous event-camera paradigm. It is shown that the contrast-threshold mechanism of the event camera ($|\Delta L| \geq C$) and the PCC-based change criterion address the same problem ("did this signal change enough?") at different granularities, and that the PCC view provides the very property the event-camera view lacks: a relational, scene-adaptive, distribution-free reference. PCC is then lifted to the three standard intermediate representations of the event stream, Time Surface (TS), Event Frame (EF), and Voxel Grid (VG), yielding three integrity and gating metrics (r-TS, $r_2$-EF, r-VG).

Throughout this work epistemic status is made explicit. The structural isomorphism (Section 4) and the algebraic constructions of the three metrics (Section 5) are derivations from published mathematics. The quantitative behavior shown in Section 6 is computed on a procedural-synthetic event stream, i.e., events generated by direct simulation of the emission model on a known scene, rather than recorded by a camera or generated from real video by a tool such as v2e [17], and is labelled [ILLUSTR.-SYNTH.] accordingly.

***Contributions.***

- **C1 — Structural isomorphism.** A formal correspondence is established between the event-emission rule $|\Delta L| \geq C$ and the PCC change criteria of [8, 9], identifying the threshold C as the static, hardware-fixed analogue of the adaptive scene-mean reference of PCC. This correspondence has not been previously articulated.
- **C2 — PCC-Event framework (PCC-TS / $r_2$-EF / PCC-VG).** Three task-agnostic, distribution-free integrity and gating metrics operating directly on the standard event representations are formalized.

- **C3 — Pipeline latency analysis.** The irreducible hardware acquisition latency (1–10 µs/event) is separated from the reducible pipeline latency, and PCC-Event is shown to provide cumulative gains on the latter without modifying the former.
- **C4 — Symmetric information-loss analysis.** A balanced account is provided of what PCC-Event loses with respect to the raw stream, and conversely.
- **C5 — XAI-by-construction for SOTIF.** The algebraic, non-parametric nature of PCC is shown to satisfy the explainability requirements of ISO/PAS 8800:2024 §6.3 without recourse to post-hoc methods.
- **C6, Procedural-synthetic illustrative analysis.** All three metrics are computed on a known synthetic stream that includes a tunnel-dip integrity event; results show that r_C cleanly separates coherent (≈ 0.93) from anomalous (< 0) regimes, illustrating the BiasBench-gap mechanism that the framework is designed to fill. A dynamic 25-second demonstration of the three metrics evolving synchronously is provided as a supplementary video.

The remainder of the paper is organized as follows. Section 2 reviews the state of the art. Section 3 establishes the formal background of event cameras and their intermediate representations. Section 4 develops the PCC bridge and the structural isomorphism (C1). Section 5 introduces the PCC-Event framework (C2). Section 6 presents the latency, information-loss, and procedural-synthetic illustrative analyses (C3, C4, C6). Section 7 discusses XAI properties under ISO/PAS 8800:2024 (C5). Section 8 positions the framework against the open scientific locks identified in the literature, and Section 9 concludes with a roadmap for empirical validation.

## 2. Related Work

### 2.1 Event-based vision: representations and learning

The comprehensive survey by Gallego et al. [1] consolidates the foundations of event-based vision: the per-pixel asynchronous emission model, the standard intermediate representations (Time Surface, Event Frame, Voxel Grid), and the principal application domains. The survey by Cimarelli et al. [2] updates this view with neuromorphic hardware deployment for embedded applications. Gehrig and Scaramuzza [3] established a milestone result, demonstrating sub-millisecond end-to-end latency on a neuromorphic processor with hybrid event/frame fusion. The work operates at the representation layer of [1, 2]: instead of proposing a new event representation, the work introduces a new algebraic metric defined on the existing ones, designed for stream integrity rather than for a downstream perception task.

### 2.2 Stream-quality metrics and the BiasBench gap

The contrast threshold C of an event camera is governed by hardware bias registers. Per-pixel manufacturing imperfections induce hot pixels (effective C too low, continuous noise) and dead pixels (effective C too high, under-detection). Ziegler et al. [6], in BiasBench (CVPR 2025), introduced the first benchmark for event-camera bias calibration and explicitly state that no metric operating directly on the event stream, without dependence on a downstream task, currently exists. Their evaluation uses detection-accuracy curves as proxies, which is the dependence that SOTIF certification cannot accept. PCC-Event is designed to fill this gap by construction.

## 2.3 Adaptive temporal and spatial segmentation

SpikeSlicer [7] (NeurIPS 2024) is the most direct technical neighbor: it learns when to segment the event stream temporally, using an ANN-based feedback loop. The training signal of SpikeSlicer, however, is a tracking or recognition loss; that is, it depends on a downstream applicative task. PCC-Event differs in two ways: (a) it segments where as well as when (spatial ROIs via $r_2$-EF in addition to temporal gating via r-VG), and (b) its supervisory signal r-TS is algebraic and applicative-task-free, a property required for SOTIF/ISO 21448 compliance.

## 2.4 Safety standards and explainable perception

ISO 21448 (SOTIF) [4] addresses functional insufficiencies in ADS, particularly performance limitations of perception components under unknown but reasonably foreseeable scenarios. ISO/PAS 8800:2024 [5] is the first standard explicitly targeting AI-component safety in automotive applications and §6.3 imposes traceable, interpretable evidence of model behavior. Recent surveys [10, 11] highlight that ISO 26262 was not designed for data-driven AI components and that ISO 21448 lacks quantitative acceptance criteria. The systematic review on explainable AI for trustworthy autonomous driving [12] identifies the absence of algebraic, model-free explainability primitives as a recurrent shortcoming of post-hoc XAI methods. PCC, by virtue of its non-parametric algebraic definition, sidesteps the post-hoc requirement entirely.

## 2.5 PCC in computer vision

PCC and its normalized cross-correlation (NCC) variant have a long history in computer vision for template matching, image registration, and stereo correspondence. The recent survey by Ghosh and Gallego [13] documents the long-standing use of correlation-based criteria, including normalized cross-correlation, among the families of techniques applied to event-based stereo matching, but no unified integrity framework derived from these criteria has been proposed for stream-quality assessment. Gallego et al. [14] introduced contrast-maximization (a philosophically related variance-based criterion) for ego-motion estimation. The two prior contributions [8, 9] established PCC as a global redundancy filter and a binarized ROI selector on frame-based images, in real-time autonomous-navigation contexts. The present paper systematizes the lift of these mechanisms from frames to event representations.

# 3. Background: Event Cameras and Intermediate Representations

## 3.1 Event-emission model [ESTABLISHED]

Each pixel at coordinates (x, y) continuously monitors the logarithm of the illuminance L(x, y, t). An event e is emitted at instant t whenever the variation exceeds the contrast threshold C:

$$\begin{aligned} &e = (x, y, t, p), \quad p = \mathrm{sign}(\Delta L(x, y, t)) \\ &\Delta L(x, y, t) = \log L(x, y, t) - \log L(x, y, t - \Delta t) \\ &\text{Condition: } |\Delta L(x, y, t)| \geq C, \quad C > 0 \end{aligned} \tag{1}$$

The output is an asynchronous time-ordered sequence E = {e_k = (x_k, y_k, t_k, p_k)}. The threshold C is a global scalar fixed by hardware bias-register configuration. Per-pixel manufacturing variability produces hot and dead pixels and constitutes the open lock identified by [6].

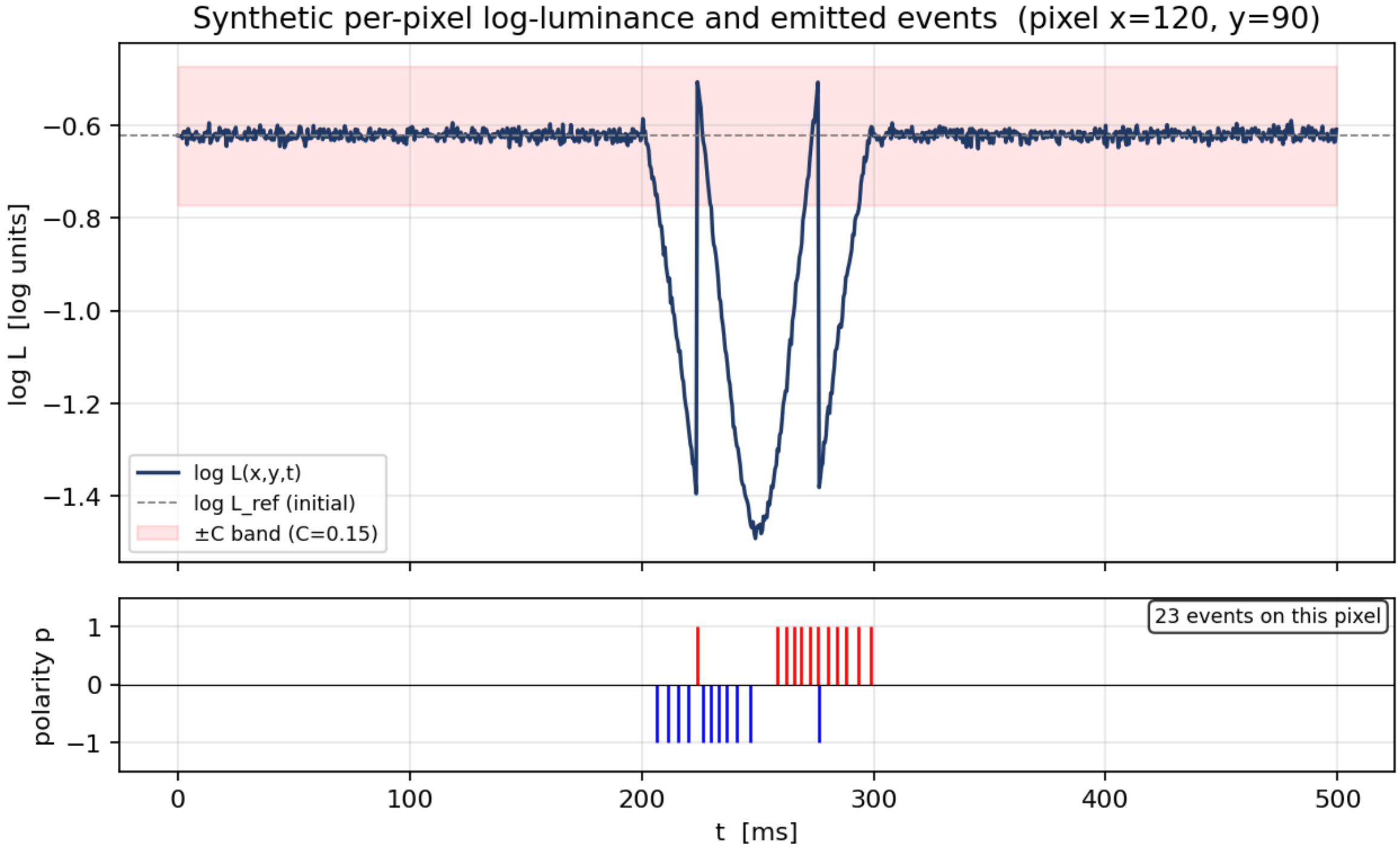

*Fig. 1. [ILLUSTR.-SYNTH.] Single-pixel illustration of the event-emission model (Eq. 1) on the procedural-synthetic stream defined in §6.1. Top: log-luminance L(x, y, t) at pixel (120, 90), with the ±C band around the initial reference (C = 0.15). The synthetic tunnel-dip scenario between 200 and 300 ms causes log L to exit the band downwards (negative events) and then upwards (positive events). Bottom: the 23 events emitted on this pixel, each at the time step where the band was crossed. The reference is updated after each event, in line with the standard event-camera firmware.*

## 3.2 Standard intermediate representations [ESTABLISHED]

Three representations convert the asynchronous stream into dense tensors processable by classical algorithms.

**Time Surface (TS).** For each pixel, the value encodes the recency of the last event:

$$TS(x, y, t) = \exp(-(t - \tau(x, y)) / \delta), \quad TS \in [0, 1] \tag{2}$$

where $\tau(x, y)$ is the timestamp of the last event at pixel (x, y) and $\delta > 0$ is a temporal-decay constant.

**Voxel Grid (VG).** Discretization of the stream into a 3D tensor by accumulation:

$$VG(x, y, b) = \Sigma_k\, p_k \cdot \kappa(b - b_k), \quad b_k = B \cdot (t_k - t_0) / (t_N - t_0) \tag{3}$$

with $\kappa(\cdot)$ a bilinear or Gaussian kernel and B the number of temporal bins.

**Event Frame (EF).** Accumulation of polarities per pixel over a temporal window $[t_0, t_0 + T]$:

$$EF(x, y) = \Sigma_\{k : t_k \in [t_0, t_0+T]\}\, p_k \cdot \delta(x - x_k, y - y_k) \tag{4}$$

EF takes signed integer values: the net polarity flux per pixel.

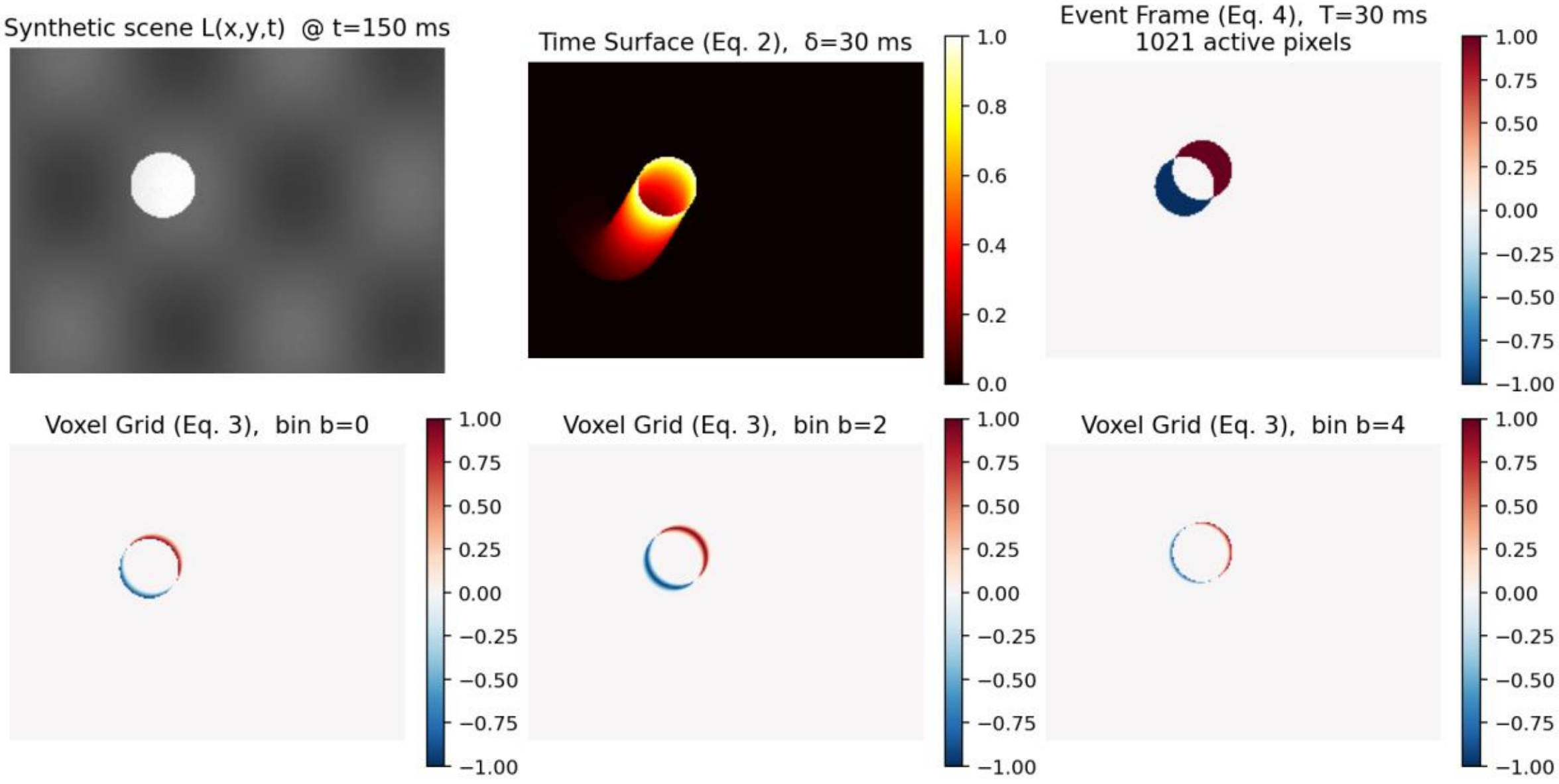

*Fig. 2. [ILLUSTR.-SYNTH.] The three-standard event-stream representations (Eqs. 2–4) computed on the procedural-synthetic stream at t = 150 ms with window T = 30 ms. Top row, left to right: ground-truth scene L(x, y, t) (a moving disk over a textured background); Time Surface (Eq. 2) showing the typical comet-trail pattern of recent events; Event Frame (Eq. 4) with 1,021 active pixels across the leading and trailing edges of the disk in motion (red = positive polarity, blue = negative). Bottom row: three temporal bins of the Voxel Grid (Eq. 3), illustrating the temporal decomposition of the same window into B = 5 bins.*

## 4. The PCC Bridge: Structural Isomorphism with the Event Paradigm

This section establishes contribution C1: a formal correspondence between the contrast-threshold mechanism of the event camera and the PCC change criteria used in prior work [8, 9]. The correspondence is not rhetorical; it is mathematically grounded and exposes the precise asymmetry that the PCC-Event framework will exploit.

### 4.1 Pearson correlation coefficient [ESTABLISHED]

For two vectors X = (x_1, ..., x_n) and Y = (y_1, ..., y_n), the Pearson correlation coefficient is

$$r(X, Y) = \Sigma_i (x_i - \bar{x})(y_i - \bar{y}) \,/\, \sqrt{[\, \Sigma_i (x_i - \bar{x})^2 \cdot \Sigma_i (y_i - \bar{y})^2 \,]}, \quad r \in [-1, +1] \qquad \textbf{(5)}$$

The fundamental property for safety-of-functionality (SdF) is that PCC is invariant under translation and scaling of X and Y, and presupposes no a-priori noise distribution. This non-parametric property is the one required by SOTIF certification for operational design domains where the noise distribution cannot be assumed [4].

### 4.2 First milestone: global change criterion [ESTABLISHED]

In Miranda Neto et al. [8], PCC is computed between the current frame I_t and a reference frame I_ref over a spatial window Ω. The decision rule is

$$
\begin{aligned}
&r_global(t) = r( I_t|_\Omega , I_ref|_\Omega ) \\
&Decision: \; process\ I_t \;\; if \;\; r_global(t) < \theta \\
&\qquad discard\ I_t \;\; if \;\; r_global(t) \geq \theta
\end{aligned} \quad \textbf{(6)}
$$

with $\theta \in [0, 1]$ a pre-selected redundancy threshold. The objective is to reduce real-time computational load by activating the processing pipeline only when visual information has changed sufficiently, without an a-priori noise model.

### 4.3 Second milestone: pixel-level binarized ROI selection [ESTABLISHED]

In Miranda Neto et al. [9], PCC is applied in two successive stages. The first computes the global means $\bar{x}$ and $\bar{y}$. The second applies pixel-wise binarization, and here the algebraic property is essential: the result is constrained to take exactly the values $\{-1, +1\}$ because numerator and denominator share the same structure up to sign:

$$
\begin{aligned}
&r_2(x_i, y_i) = (x_i - \bar{x})(y_i - \bar{y}) / ( |x_i - \bar{x}| \cdot |y_i - \bar{y}| ) \\
&\qquad = sign[ (x_i - \bar{x})(y_i - \bar{y}) ] \;\; \in \{-1, +1\} \\
&ROI = \{ (x_i, y_i) : r_2 = -1 \}
\end{aligned} \quad \textbf{(7)}
$$

The critical computational property [ESTABLISHED] is that $r_2$ requires no costly multiplication and no square root, only additions (for $\bar{x}$, $\bar{y}$) and sign comparisons. Complexity: $O(N)$ with $N = |\Omega|$. This enables pure-integer FPGA or NPU implementations in $O(N)$ clock cycles, a property that will prove crucial for the lift to the event domain (Section 6).

### 4.4 Structural isomorphism [SOLID]

The conjoint analysis of [8] and [9] reveals that those works addressed the same problem that the event camera (DVS, Prophesee) solves in hardware since 2008 [1], with a difference in granularity, not in philosophy. Figure 3 visualizes the correspondence. Table 1 makes it explicit dimension by dimension.

| Dimension | CCA 2007 / IROS 2011 [8, 9] | Event camera (DVS / Prophesee) [1] |
|---|---|---|
| Central question | Has this frame/pixel changed enough? | Has this pixel changed enough? |
| Reference | μ(scene), adaptive (IROS 2011) / I_ref (CCA 2007) | L(x, y, t − Δt), per-pixel stored level |
| Criterion | $r_2 = -1 \rightarrow$ ROI ; $r_global < \theta \rightarrow$ process | $\lvert \Delta L(x, y, t) \rvert \geq C \rightarrow$ emit event |
| Granularity | Frame (CCA 2007) / pixel vs. μ (IROS 2011) | Per-pixel, asynchronous |
| Adaptive reference? | Yes, recomputed at each frame, non-parametric | No, C is fixed by hardware bias |
| Goal | Eliminate temporal redundancy | Eliminate temporal redundancy |

*Table 1. Structural correspondence between PCC-based change criteria [8, 9] and the event-camera emission model [1].*

The formal correspondence is summarized as:

$$\begin{aligned}
&\textit{discard } I_t \quad \text{if} \quad r_global(t) \geq \theta \\
&\textit{Event:} \quad \textit{emit } e \quad \textit{iff} \quad |\Delta L(x, y, t)| \geq C \\
&\textit{Analogy:} \quad \theta \leftrightarrow C, \quad r_global(t) \leftrightarrow 1 - |\Delta L|/L \\
&\textit{ROI} \quad \text{if} \quad r_2(x, y) = -1 \quad \Leftrightarrow \quad I(x, y) < \mu \\
&\textit{Event:} \quad \textit{event } p = -1 \quad \text{if} \quad \Delta L(x, y, t) < -C
\end{aligned} \tag{8}$$

The critical asymmetry [SOLID]: PCC uses a relational, scene-adaptive reference (the running scene mean μ, recomputed at each frame), whereas the event camera uses an absolute reference C fixed by hardware. This asymmetry founds the original contribution of PCC-Event: it provides a relational, distribution-free reference where the hardware paradigm instead provides an absolute fixed one. The two are complementary rather than competing.

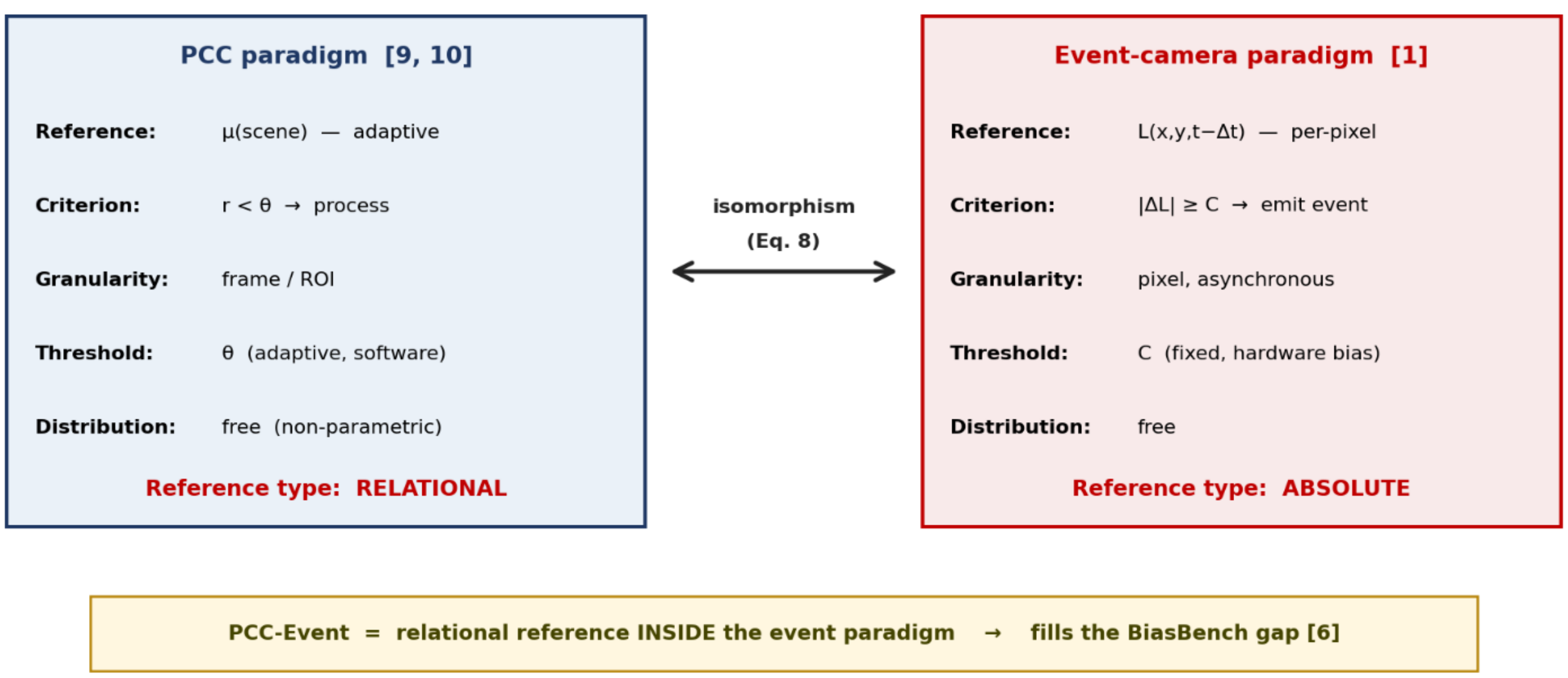

*Fig. 3. Structural isomorphism between the PCC-based change paradigm of [8, 9] and the event-camera emission paradigm of [1]. Both paradigms answer the same question, has this signal changed enough? — with non-parametric, distribution-free criteria. They differ in the nature of the reference (relational and scene-adaptive vs. absolute and hardware-fixed) and in the granularity (frame/ROI vs. asynchronous pixel). PCC-Event introduces a relational reference inside the event paradigm, filling the BiasBench gap [6].*

## 5. PCC-Event Framework

This section presents contribution C2: three task-agnostic algebraic metrics, PCC-TS, $r_2$-EF, PCC-VG, that lift the mechanisms of (6)–(7) to the event representations of (2)–(4). The classification is [SOLID]: the constructions are well-founded extensions of [8, 9] supported by precedent in [13, 14], and their behavior is illustrated on the procedural-synthetic stream of §6.1.

### 5.1 PCC-TS: structural coherence on Time Surfaces [SOLID]

Let ROI_A and ROI_B be two regions of the pixel grid of dimension $H_R \times W_R$. The structural-coherence metric between Time Surfaces is

$$r_{TS}(A, B, t) = r(\, vec(TS_A(t)) \,,\, vec(TS_B(t)) \,) \quad \in [-1, +1] \tag{9}$$

with vec(TS_A(t)) the vectorization of the Time Surface restricted to A. A value $r_{TS} \approx +1$ indicates structurally coherent temporal activity between the two ROIs; $r_{TS} \ll 1$ signals incoherence, a candidate symptom of threshold-C drift or spurious-event injection.

For threshold-C monitoring, the observed Time Surface is compared to one predicted by IMU-integrated ego-motion:

$$\begin{aligned} &r_C(t) = r(\, vec(TS_{obs}(t)) \,,\, vec(TS_{pred}(t)) \,) \quad \in [-1, +1] \\ &r_C(t) \geq \theta_{high} \;\rightarrow\; \text{flow coherent, C plausible} \\ &r_C(t) < \theta_{low} \;\rightarrow\; \text{ALARM: C likely mis-calibrated} \end{aligned} \tag{10}$$

**Differentiation from BiasBench [SOLID].** BiasBench [6] uses detection performance (DET curves) as a quality proxy. The metric $r_C$ in (10) is computed directly from the stream, without invoking any applicative task, the property required by SOTIF [4] and ISO/PAS 8800 §6.3 [5]. The illustrative behavior of (10) on the synthetic tunnel-dip scenario is shown in Figure 6.

### 5.2 $r_2$-EF: ROI selection on Event Frames [SOLID]

The binarization (7) of [9] is applied to the Event Frame, restricting to active pixels to avoid mean degeneracy:

$$\begin{aligned} &\text{Step 1:}\;\; \Omega_{act} = \{\, (x, y) : EF(x, y) \neq 0 \,\} \\ &\qquad \mu_{EF} = (1/|\Omega_{act}|) \cdot \Sigma_{(x,y)\in\Omega_{act}} EF(x, y) \\ &\text{Step 2:}\;\; r_2_EF(x_i, y_i) = sign(\, EF(x_i, y_i) - \mu_{EF} \,) \quad \in \{-1, +1\} \\ &\qquad ROI_{EF} = \{\, (x_i, y_i) \in \Omega_{act} : r_2_EF = -1 \,\} \end{aligned} \tag{11}$$

The restriction to $\Omega_{act}$ is essential [ESTABLISHED]. Without it, the $EF(x, y) = 0$ pixels, a majority in any sparse stream, would dominate $\mu_{EF} \approx 0$ and $r_2$_EF would be negative for almost all active pixels, destroying the discriminative power. The formulation $r_2_EF = sign(EF - \mu_{EF})$ is the exact transposition of the $r_2$ mechanism to event frames: zero multiplications, zero square roots, pure integer arithmetic. The mechanism is illustrated in Figure 4.

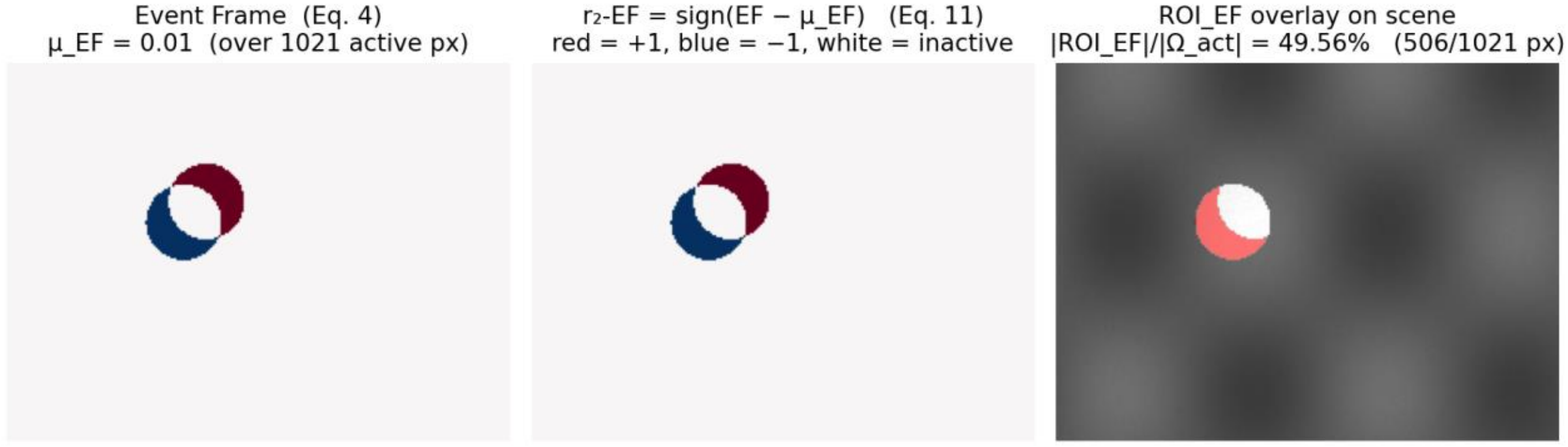

*Fig. 4. [ILLUSTR.-SYNTH.] $r_2$-EF mechanism (Eq. 11) on a synthetic Event Frame at t = 150 ms (T = 30 ms). Left: input Event Frame, μ_EF ≈ 0.01 over 1,021 active pixels. Centre: $r_2$_EF = sign(EF − μ_EF) ∈ {−1, +1} (red = +1, blue = −1, white = inactive). Right: ROI_EF (red overlay) on the underlying scene; |ROI_EF|/|Ω_act| = 49.6 % in this single window. The selection is computed with O(|Ω_act|) integer comparisons after the initial mean — no multiplications, no square roots.*

### 5.3 PCC-VG: temporal coherence on Voxel Grids [SOLID]

The redundancy criterion (6) lifts naturally to consecutive Voxel Grids:

$$r_VG(t, t+\Delta) = r(\, vec(VG(t)) \,,\, vec(VG(t+\Delta)) \,) \quad \in [-1, +1]$$
$$r_VG \geq \theta \;\rightarrow\; \text{flow stable, pipeline trigger NOT required}$$
$$r_VG < \theta \;\rightarrow\; \text{flow changing, trigger downstream pipeline} \qquad \textbf{(12)}$$

**Precedents [SOLID].** (i) Gallego et al. [14] maximize the variance of warped events (contrast-maximization) — a philosophy related to PCC; (ii) the survey by Ghosh and Gallego [14] documents normalized cross-correlation among the criteria used for event-based stereo matching, and NCC is formally identical to PCC when the signals are centered. The PCC-VG mechanism extends this lineage from a registration tool to a temporal-redundancy gating criterion. The illustrative gating behavior on the synthetic stream is shown in Figure 5.

## 6. Latency, Information-Loss, and Procedural-Synthetic Illustrative Analyses

### 6.1 Procedural-synthetic illustrative stream

To make the behavior of the three metrics observable without resorting to fabricated experimental data, a controlled synthetic event stream is generated by direct simulation of (1). The scene is a 240 × 180 sensor with: (a) a static low-frequency textured background; (b) a moving disk translating across the field of view with mild vertical oscillation; (c) a synthetic tunnel-dip episode in which the global illumination drops by ≈ 85 % between t = 200 ms and t = 300 ms. Luminance is integrated at a 100 μs simulation tick; events are emitted whenever |log L − log L_ref| ≥ C with C = 0.15, and the per-pixel reference is updated upon emission, faithful to standard event-camera firmware. Mild per-pixel noise (1 % multiplicative) is injected to avoid degenerate determinism. Over 500 ms of simulated time the scene produces 893,793 events. It is emphasized that this is procedural simulation. Quantitative results obtained from this stream are labelled [ILLUSTR.-SYNTH.] throughout.

### 6.2 Two latency levels [ESTABLISHED]

A rigorous analysis must distinguish two fundamentally different levels of latency.

- **Hardware-acquisition latency (1–10 µs, software-irreducible):** the delay between a luminance change and the emission of (x, y, t, p). Determined by the physics of the neuromorphic sensor; no software treatment can reduce it.
- **Pipeline latency (architecture-reducible):** the delay between event reception and a perception decision. Gehrig and Scaramuzza [3] report < 1 ms end-to-end on dedicated NPUs, but this concerns the processing of all received events.

PCC-Event offers no gain at the hardware level. Its gains operate strictly at the pipeline level, through two complementary mechanisms inherited from [8, 9].

### 6.3 Pipeline-latency reduction mechanisms [SOLID]

**(a) Trigger-rate reduction via PCC-VG.** The criterion $r_VG(t, t+\Delta) < \theta$ activates the downstream pipeline. The fraction f_VG of consecutive windows whose r_VG exceeds the threshold (i.e., that can be safely discarded) is scene-dependent: high for scenes with extended quiescent regimes, near zero for scenes in continuous motion. On the synthetic stream of §6.1, which is dominated by continuous disk motion and a sharp tunnel transition, f_VG ≈ 9 % [ILLUSTR.-SYNTH., Fig. 5]. On real urban scenes with stop-and-go traffic the literature suggests substantially higher values.

**(b) Spatial-load reduction via $r_2$-EF.** The criterion $r_2_EF = \mathrm{sign}(EF - \mu_EF)$ requires $O(|\Omega_act|)$ integer comparisons. On the same synthetic window, $|ROI_EF|/|\Omega_act| = 49.6$ % [ILLUSTR.-SYNTH., Fig. 4], a moderately selective regime, with the selection corresponding visually to the trailing (negative-polarity) edge of the moving disk. Scenes with stronger contrast separation between leading and trailing edges typically yield more selective ROI_EF.

The cumulative gain is, to first order,

$$Charge_downstream / Charge_total \approx (1 - f_VG) \cdot |ROI_EF| / |\Omega_act| \quad \textbf{(13)}$$

On the synthetic stream the product evaluates to ≈ 0.91 × 0.50 ≈ 45 % [ILLUSTR.-SYNTH.]. Real datasets with quieter dynamics will likely produce smaller fractions; aggressive scenes (highway at night, rain) larger ones. The point of (13) is the structure, not a universal constant.

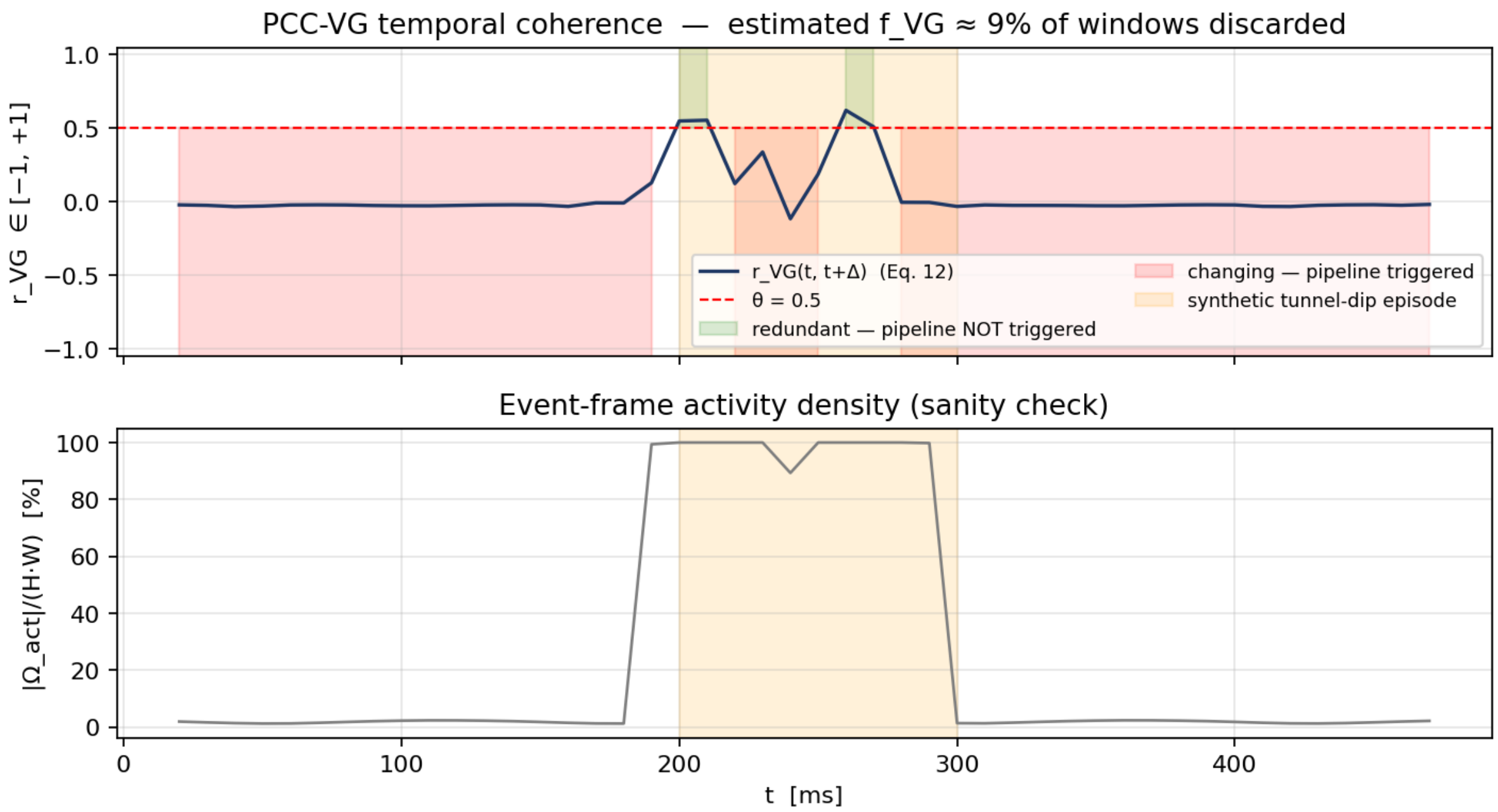

*Fig. 5. [ILLUSTR.-SYNTH.] PCC-VG temporal-coherence gating (Eq. 12) on the synthetic stream. Top: r_VG between consecutive Voxel-Grid windows (Δ = 20 ms). Green-shaded regions (r_VG ≥ θ = 0.5) correspond to redundant windows for which the downstream pipeline does not need to be triggered; red-shaded regions are the complement. Orange band: synthetic tunnel-dip episode. Bottom (sanity check): event-frame activity density |Ω_act|/(H·W) over time, showing that the dip episode produces an order-of-magnitude increase in emission rate as the global illumination crosses the per-pixel ±C bands. On this scene, dominated by continuous disk motion, only ≈ 9 % of windows are gated as redundant; on quasi-static urban scenes the fraction is expected to be substantially higher.*

## 6.4 Pipeline-comparison summary

| Pipeline stage | Event camera (state of the art) [1, 3] | PCC-Event (proposed) |
|---|---|---|
| Hardware acquisition | 1–10 µs/event, irreducible by software. | Identical. PCC-Event operates downstream with the same sensor. |
| Decision (when?) | All events enter the pipeline; no intrinsic pre-filter. | PCC-VG: r_VG < θ → trigger. f_VG ≈ 9 % on synthetic [ILLUSTR.-SYNTH.]. |
| Spatial selection (where?) | Static hardware ROI or dense processing. | $r_2$-EF: O(\|Ω_act\|) integer comparisons. \|ROI_EF\|/\|Ω_act\| ≈ 50 % on synthetic [ILLUSTR.-SYNTH., Fig. 4]. |
| Stream-quality metric | None intrinsic. Hot/dead pixels undetectable without downstream task [6]. | PCC-TS: r_C(t) = r(TS_obs, TS_pred). r_C drops from 0.93 (coherent) to < 0 (alarm) on synthetic dip [ILLUSTR.-SYNTH., Fig. 6]. |
| Certifiable downstream | < 1 ms end-to-end on full event stream [3]. | < 1 ms on ROI_EF only, combined PCC-VG + $r_2$-EF charge, see Eq. 13. |

*Table 2. Pipeline-latency comparison: event camera alone vs. PCC-Event.*

### 6.5 Failure modes [SOLID]

- **Critical case 1, High-density scene (highway, rain, night):** if $|\Omega_act|/(H \cdot W) \to 1$, the selectivity of $r_2$-EF degrades ($\mu_EF \to 0$). PCC-EF gain may be negligible.
- **Critical case 2, Sudden perturbation of a static scene:** PCC-VG may have discarded the window preceding a critical event ($r_VG \geq \theta$ on the static window) — late trigger.

These failure modes do not invalidate the framework. The PCC-Event contributions are orthogonal to raw-stream latency: integrity (r_C), gating (r_VG), and spatial selection ($r_2$_EF) are three contributions absent from the raw stream. If pipeline gain is modest, the SOTIF integrity contribution remains intact, as the synthetic dip experiment of Figure 6 illustrates directly.

### 6.6 PCC-TS integrity monitor, illustrative behavior

Figure 6 shows the behavior of the PCC-TS integrity monitor (Eq. 10) on the synthetic stream. The 'predicted' Time Surface is computed from a parallel simulation of the same scene without the tunnel dip, i.e., the prediction the system would produce given knowledge of ego-motion alone. Outside the dip episode, observed and predicted Time Surfaces are nearly identical and $r_C(t) \approx 0.93$. During the dip, the surge of additional events caused by the global illumination drop pushes the observed Time Surface away from the prediction, and r_C(t) collapses below zero. The metric crosses $\theta_low = 0.4$ cleanly at the onset and offset of the dip. Crucially, the alarm is computed without reference to any applicative task, it depends only on the algebraic relation between two stream-derived quantities, satisfying the BiasBench gap criterion [6].

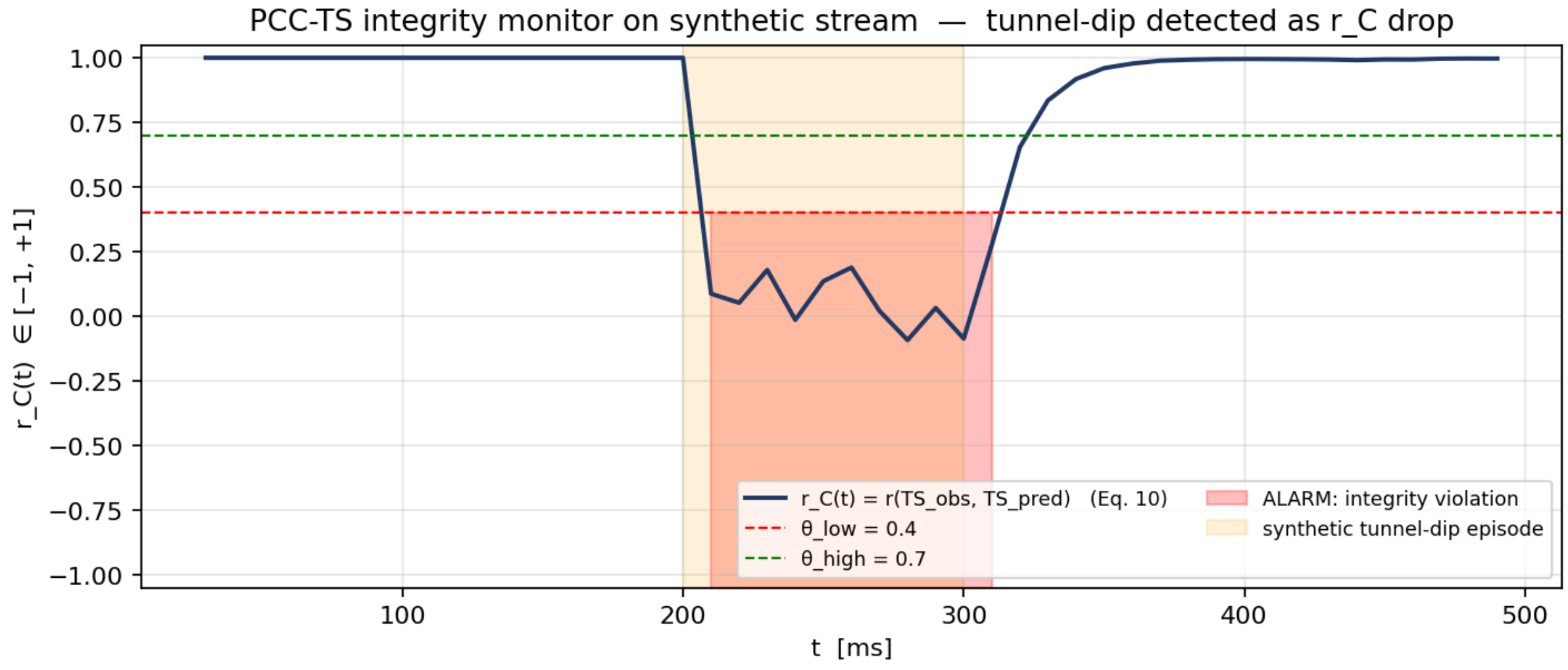

*Fig. 6. [ILLUSTR.-SYNTH.] PCC-TS integrity monitor r_C(t) (Eq. 10) on the synthetic stream. Outside the tunnel-dip episode (orange band, 200–300 ms), $r_C(t) \approx 0.93$, indicating that the observed event flow matches the IMU-predicted flow. During the dip, r_C drops below zero and triggers the integrity alarm (red band) by crossing $\theta_low = 0.4$. The metric returns to ≈ 1 within ~ 30 ms after the dip ends. The supervisory signal is purely algebraic and stream-derived; no downstream applicative task is used.*

### 6.7 Symmetric information-loss analysis [ESTABLISHED for PCC side]

Every representation incurs a trade-off between complexity reduction and information preservation. The analysis must be conducted symmetrically, both what PCC-Event loses and what the raw stream loses by virtue of not having a relational metric must be reported.

| Dimension | PCC-Event loses (vs. raw stream) | Raw stream loses (vs. PCC-Event) |
|---|---|---|
| Intra-window temporal resolution | Real loss. $r_2$-EF and PCC-TS operate on windows or with constant δ; sub-μs causal relations within a window are lost. | Partial loss. Raw stream produces no structured temporal metric. PCC-TS encodes inter-event memory through exponential decay τ. |
| Per-pixel amplitude | Real loss. $r_2$-EF binarization to {−1, +1} loses event-frequency amplitude. | Real loss. Raw (x, y, t, p) does not produce relative ROI-vs.-mean comparisons without accumulation. |
| Long-range spatial correlation | Not applicable. PCC-TS computes exactly r(TS_A, TS_B) between distant ROIs, its principal contribution. | Real loss. Raw stream does not quantify structural similarity between distant ROIs without further processing. |
| Threshold-C anomaly | Not applicable. This is precisely what PCC-TS provides. | Structural loss. No intrinsic metric of the threshold C, the open lock of [6]. |
| Interpretability / XAI | Not applicable. PCC is XAI by construction, algebraic, directly interpretable. | Structural loss. SSIM, PSNR, and Grad-CAM are not directly applicable to event frames, which lack the dense intensity structure these metrics assume. |
| Embedded efficiency (FPGA/NPU) | Advantage. $r_2$-EF: pure integer arithmetic, O(\|Ω_act\|) cycles. | Neutral. Raw stream is inherently compact, but downstream SNN/CNN processing requires dedicated architectures. |

*Table 3. Symmetric information-loss analysis: PCC-Event vs. raw event stream.*

Recommended operating regimes. PCC-Event is preferable for: moderate-density scenes (urban 20–60 km/h); SOTIF / ISO/PAS 8800 certification applications; embedded systems with limited memory bandwidth. Raw event stream alone is preferable for: hard sub-millisecond latency on optimized NPUs; dense optical flow or 3D reconstruction; ultra-high-density scenes (highway at night/rain).

## 7. Explainability Under ISO/PAS 8800:2024

This section develops contribution C5: PCC-Event satisfies the explainability requirements of ISO/PAS 8800:2024 §6.3 [5] by construction, without recourse to post-hoc XAI methods. The argument rests on three properties of (5):

- **Algebraic interpretability.** $r \in [-1, +1]$ is a closed-form algebraic quantity. A score $r_C < \theta_{low}$ in a ROI constitutes a traceable, interpretable, auditable integrity alarm, not an output of an opaque model.
- **Distribution freedom.** PCC presupposes no noise distribution. SOTIF [4] explicitly requires safety arguments that hold across the ODD without distributional assumptions on hazardous scenarios.

- **Causal attribution.** When r_C(t) < θ_low, two candidate causes are immediately legible:

$$\textit{r_C(t)} < \theta\textit{_low} \rightarrow \textit{Cause 1: C miscalibrated (hot/dead pixel)} \quad \textit{Cause 2: unpredicted obstacle in ROI_k} \quad \textit{Action (SOTIF): increase epistemic-uncertainty mass for the cell.} \tag{14}$$

Compare with post-hoc methods on event data. Saliency-based explanation techniques such as Grad-CAM / Grad-CAM++ [18] and SHAP [19], when applied to event-frame classifiers, face a structural difficulty: the absence of dense intensity violates the spatial-saliency assumptions on which these methods depend. Counterfactual explanations [20] are compatible with the SOTIF framework but remain post-hoc and require a trained downstream model, precisely the dependence that PCC-Event avoids. The key claim [SOLID] is therefore: PCC-Event provides an algebraic XAI primitive that complements (not replaces) post-hoc methods at the perception-integrity layer.

## 8. Positioning Against Open Scientific Locks

The framework's contributions are summarized below against four open locks identified in the recent literature.

- **[OPEN] Lock 1 — Task-agnostic stream-integrity metric.** BiasBench [6] confirms its absence. PCC-TS (10) addresses this gap; the synthetic dip experiment of Figure 6 illustrates the mechanism.
- **[OPEN] Lock 2 — Adaptive spatial segmentation without absolute threshold.** SpikeSlicer [7] resolves temporal segmentation; spatial segmentation by scene-adaptive ROI without an absolute threshold remains an open problem. $r_2$-EF (11) constitutes an algebraic answer.
- **[OPEN] Lock 3 — Heterogeneous fusion under extreme uncertainty.** Direct fusion in event space with epistemic-uncertainty propagation (Dempster–Shafer, credal occupancy grids) is an open problem. PCC-TS provides a stream-derived evidence source compatible with such formalisms, explored as future work outside this paper's scope.
- **[OPEN] Lock 4 — Certifiability of asynchronous sensors.** ISO 26262 [21], ISO 21448 [4], and ISO/PAS 8800:2024 [5] do not yet explicitly cover neuromorphic sensors. PCC-TS, being algebraic and non-parametric, is a natural candidate for inclusion in such standardization efforts (e.g., ISO TC 22 / SC 32).

## 9. Conclusion and Roadmap

This paper has presented PCC-Event, a unified algebraic framework that lifts the Pearson Correlation Coefficient, historically used in [8, 9] for redundancy filtering and ROI selection on frame-based images, to the three standard event representations (Time Surface, Event Frame, Voxel Grid). The framework provides three task-agnostic metrics: PCC-TS for stream integrity, $r_2$-EF for adaptive ROI selection, and PCC-VG for temporal redundancy gating. The structural isomorphism between the PCC change criterion and the contrast-threshold mechanism of the event camera has been established, pipeline latency and information loss have been analyzed symmetrically, and the explainability properties of PCC under ISO/PAS 8800:2024 §6.3 have been argued. Each metric has been illustrated on a procedural-synthetic stream including a tunnel-dip integrity scenario, where r_C cleanly separates coherent regimes (r_C ≈ 0.93) from anomalous ones (r_C < 0).

This paper is intentionally a theoretical framework paper supplemented by procedural-synthetic illustrations. The empirical validation of all [SOLID] claims on real recorded data constitutes the next stage of the research program.

- **Empirical validation on DSEC [15], GEN1 [16], and N-CARS [22].** Quantification of f_VG, |ROI_EF|/|Ω_act|, and threshold-C drift detection in real recordings. Comparison with SpikeSlicer [7].
- **Lift of PCC to the raw asynchronous stream via spiking neural networks.** An SNN trained to approximate PCC directly on (x, y, t, p) on neuromorphic hardware (Loihi 2, DYNAP-CNN), supervised by an algebraic, task-agnostic, loss derived from PCC-TS.
- **Integration with credal occupancy grids.** Mapping r_C(t) → mass functions in Dempster–Shafer credal grids for multi-sensor fusion under extreme uncertainty.

By proposing a relational, distribution-free reference exactly where the event-camera paradigm provides an absolute fixed one, PCC-Event aspires to be a step toward perception modules that are fast and certifiable rather than fast or certifiable.

## Reproducibility Statement

All quantitative results in this preprint marked [ILLUSTR.-SYNTH.] are obtained from a self-contained Python simulation that (i) generates the synthetic scene by direct evaluation of the luminance field, (ii) emits events according to (1) at a 100 µs simulation tick with $C = 0.15$, and (iii) computes the Time Surface, Event Frame, Voxel Grid, and the three PCC-Event metrics on the resulting stream. The simulation produces 893,793 events over 500 ms on a 240 × 180 sensor.

**Code and figures.** The full simulation script, the figure-generation pipeline, and the rendered figures are released alongside this paper at git repository. The repository contains: simulate.py (the 6 figures), build_video.py (the demonstration video), all 6 figure PNGs in figures/, the rendered MP4 in video/, and a README.md with reproduction instructions.

https://github.com/ademiran/Task_Agnostic_Algebraic_Integrity_Metric_for-Event_Camera_Streams.git

**Supplementary video.** A 25-second demonstration video (1280×720, 24 fps) showing the synchronous evolution of the synthetic scene, the three event representations (TS, EF, VG), the dynamic ROI selected by $r_2$-*EF*, and the rolling integrity and gating curves *r_C(t)* and *r_VG(t)* is available at: https://youtu.be/nwIegaTOcrU. The integrity-alarm overlay activates automatically during the synthetic tunnel-dip episode, demonstrating in real time the BiasBench-gap mechanism that the framework is designed to fill.

## Acknowledgements

The author thanks the institutional support of the Federal University of Lavras (UFLA), Brazil.

## References

**[1]** G. Gallego, T. Delbrück, G. Orchard, C. Bartolozzi, B. Taba, A. Censi, S. Leutenegger, A. J. Davison, J. Conradt, K. Daniilidis, and D. Scaramuzza, "Event-based vision: A survey," IEEE Trans. Pattern Anal. Mach. Intell., vol. 44, no. 1, pp. 154–180, 2022.

**[2]** C. Cimarelli, J. A. Millan-Romera, H. Voos, and J. L. Sanchez-Lopez, "Hardware, algorithms, and applications of the neuromorphic vision sensor: A review," Sensors, vol. 25, no. 19, art. 6208, 2025. arXiv:2504.08588.

**[3]** D. Gehrig and D. Scaramuzza, "Low-latency automotive vision with event cameras," Nature, vol. 629, pp. 1034–1040, 2024.

**[4]** ISO 21448:2022, Road vehicles — Safety of the intended functionality (SOTIF). International Organization for Standardization, 2022.

**[5]** ISO/PAS 8800:2024, Road vehicles — Safety and artificial intelligence. International Organization for Standardization, 2024.

**[6]** A. Ziegler, D. Joseph, T. Gossard, E. Moldovan, and A. Zell, "BiasBench: A reproducible benchmark for tuning the biases of event cameras," in Proc. IEEE/CVF Conf. Comput. Vis. Pattern Recognit. Workshops (CVPRW), 2025, pp. 4994–5003.

**[7]** J. Cao, M. Sun, Z. Wang, H. Cheng, Q. Zhang, S. Zhou, and R. Xu, "Spiking neural network as adaptive event stream slicer," in Adv. Neural Inf. Process. Syst. (NeurIPS), 2024. arXiv:2410.02249.

**[8]** A. Miranda Neto, L. Rittner, N. Leite, D. E. Zampieri, A. R. Lotufo, and A. Mendeleck, "Pearson's correlation coefficient for discarding redundant information in real-time autonomous navigation system," in Proc. IEEE Int. Conf. Control Applications (CCA-MSC), Singapore, 2007, pp. 426–431.

**[9]** A. Miranda Neto, A. C. Victorino, I. Fantoni, and D. E. Zampieri, "Automatic regions-of-interest selection based on Pearson's correlation coefficient," in Proc. IEEE/RSJ IROS Workshop on Visual Control of Mobile Robots, San Francisco, USA, 2011, pp. 33–38.

**[10]** H. Wang, W. Shao, C. Sun, K. Yang, D. Cao, and J. Li, "A survey on an emerging safety challenge for autonomous vehicles: Safety of the intended functionality," Engineering, vol. 33, pp. 17–34, 2024.

**[11]** R. Yu, C. Wang, Y. Zhang, and F. Zhao, "Decomposition and quantification of SOTIF requirements for perception systems of autonomous vehicles," IEEE Trans. Intell. Transp. Syst., 2025. arXiv:2501.10097.

**[12]** A. Kuznietsov, B. Gyevnar, C. Wang, S. Peters, and S. V. Albrecht, "Explainable AI for safe and trustworthy autonomous driving: A systematic review," arXiv:2402.10086, 2024.

**[13]** S. Ghosh and G. Gallego, "Event-based stereo depth estimation: A survey," IEEE Trans. Pattern Anal. Mach. Intell., 2025. arXiv:2409.17680.

**[14]** G. Gallego, M. Gehrig, and D. Scaramuzza, "Focus is all you need: Loss functions for event-based vision," in Proc. IEEE/CVF Conf. Comput. Vis. Pattern Recognit. (CVPR), 2019, pp. 12280–12289.

**[15]** M. Gehrig, W. Aarents, D. Gehrig, and D. Scaramuzza, "DSEC: A stereo event camera dataset for driving scenarios," IEEE Robot. Autom. Lett., vol. 6, no. 3, pp. 4947–4954, 2021.

**[16]** P. de Tournemire, D. Nitti, E. Perot, D. Migliore, and A. Sironi, "A large scale event-based detection dataset for automotive (GEN1)," arXiv:2001.08499, 2020.

**[17]** Y. Hu, S.-C. Liu, and T. Delbrück, "v2e: From video frames to realistic DVS events," in Proc. IEEE/CVF Conf. Comput. Vis. Pattern Recognit. Workshops (CVPRW), 2021.

**[18]** R. R. Selvaraju, M. Cogswell, A. Das, R. Vedantam, D. Parikh, and D. Batra, "Grad-CAM: Visual explanations from deep networks via gradient-based localization," in Proc. IEEE Int. Conf. Comput. Vis. (ICCV), 2017, pp. 618–626.

**[19]** S. M. Lundberg and S.-I. Lee, "A unified approach to interpreting model predictions," in Adv. Neural Inf. Process. Syst. (NeurIPS), 2017.

**[20]** A. Samadi, A. Shirian, K. Koufos, K. Debattista, and M. Dianati, "SAFE: Saliency-aware counterfactual explanations for DNN-based automated driving systems," in Proc. 26th IEEE Int. Conf. Intell. Transp. Syst. (ITSC), 2023, pp. 5655–5662.

**[21]** ISO 26262:2018, Road vehicles — Functional safety. International Organization for Standardization, 2018.

**[22]** A. Sironi, M. Brambilla, N. Bourdis, X. Lagorce, and R. Benosman, "HATS: Histograms of averaged time surfaces for robust event-based object classification (N-CARS dataset)," in Proc. IEEE/CVF Conf. Comput. Vis. Pattern Recognit. (CVPR), 2018.